\def\parder #1;#2;{{\partial #1\over \partial #2}}

\newcommand\les{\lesssim}
\newcommand\Mo {$M_{\odot}\,$}

\newcommand\refind{\hbox to 0.5 truein {\hrulefill .}}
\newcommand\beq{\begin{equation}}
\newcommand\eeq{\end{equation}}

\documentstyle[12pt, aaspp4]{article}
\slugcomment{To Appear in \apj, December 10, 1996}
\lefthead{Canuto et al.}
\righthead{convection}
\begin{document}
\title{Stellar Turbulent Convection: A Selfconsistent Model}

\author{V. M. Canuto \altaffilmark{1}, I. Goldman\altaffilmark{2,3,4} \ and
I. Mazzitelli \altaffilmark{5}}

\altaffiltext{1}{NASA Goddard Institute for Space Studies, 2880 Broadway, 
New York, NY 10025, USA, email: acvmc@nasagiss.giss.nasa.gov}
\altaffiltext{2}{School of Physics and Astronomy, Sackler Faculty of Exact
  Sciences, Tel Aviv University, Tel Aviv 69978, Israel, email:  
goldman@plato.tau.ac.il} 
\altaffiltext{3}{Visiting Scientist, Department of Condensed-Matter 
Physics, Faculty of Physics, Weizmann Institute of Science, Rehovot 76100, 
Israel} 
\altaffiltext{4}{Visiting Scientist, Harvard--Smithsonian Center For 
Astrophysics, 60 Garden Street, Cambridge, MA 02138, USA}
\altaffiltext{5}{Istituto di Astrofisica Spaziale, CNR, CP 67, 00044 
Frascati, Italy, email: aton@hyperion.ias.fra.cnr.it}

 \begin{abstract} 
We present a selfconsistent model for stellar turbulent
convection which is similar in spirit to the CM model
(Canuto \& Mazzitelli 1991) since it accounts for the full spectrum of the
turbulent eddies rather than only one eddy, as done in the mixing length 
theory
(MLT).  The model differs from the CM model in
the treatment of the rate of energy input $n_s(k)$ from the source that
generates the turbulence. In the present model, $n_s(k)$ is controlled by
both the {\it source} and {\it the turbulence} it ultimately generates,
thus ensuring a selfconsistent modeling of the turbulence. This improves
the CM model in which $n_s(k)$ was taken to be equal to the growth rate of
the {\it linear} unstable convective modes. 

However, since the  formulation of a selfconsistent treatment
 is far from simple, we were forced to use  a 
representation of
the nonlinear interactions less complete 
than the one in the CM
model. The ensuing equations were solved numerically for a wide range of
convective efficiencies. The results are the convective flux, the mean
square turbulent velocity, the root mean squared turbulent pressure and
the turbulent viscosity. 

We implemented the model in the ATON stellar
structure code and computed the evolution of a solar model. The results
are generally similar to those of the CM model and thus quite different from
the MLT. The present model requires a smaller overshoot into the
upper radiative zone than does the CM model, in accord with recent
empirical estimates. Application to POP II stars and
comparison with the very metal-poor globular cluster M68 yields an age in
the range $11\div 12$ Gyr. This is somewhat younger than the CM age, which
in turn is younger than the corresponding MLT age, a result of 
possible cosmological interest.

 \end{abstract}

\section{Introduction}

Recently, Canuto \& Mazzitelli (1991, CM) proposed an improved model for
stellar convection. Being derived from a turbulence model, it
takes into account the contribution of the full spectrum of the turbulent
convective eddies, to the convective flux. In stellar interiors the
microscopic viscosity is very small compared to the turbulent viscosity,
implying that the turbulent spectrum spans many decades in wavenumber
space. Therefore, in this respect, the CM model represents a significant
improvement over the mixing length theory approach (MLT) which is a one
eddy (the largest) approximation to the spectrum. Moreover, in the CM
model, the turbulent mixing length scale is the depth $z$, so there is no
need for an adjustable free parameter like the MLT $\alpha$-parameter. 
The resulting convective fluxes are higher than those of the MLT for high
convective efficiencies, and smaller than them for low efficiencies. The
model performs better than the MLT when applied to stellar structure
(D'Antona \& Mazzitelli, 1994; Mazzitelli, D'Antona \& Caloi 1995,
Stothers \& Chin 1995, Althaus \& Benvenuto 1996), helioseismology
(Baturin \& Mironova 1995, Monteiro et al. 1995, Antia \& Basu 1995) and
stellar atmospheres (Kupka 1995).

 In the CM model, the turbulence spectral function $E(k)$ is determined by
the timescale controlling the energy input from {\it buoyancy}, that is, the 
{\it source} that generates the turbulence. This timescale
 is expected to depend on the parameters of the source as well as on the
{\it turbulence spectrum} itself. The quantification of the latter 
dependence, within the CM model, is
far from obvious. Thus, CM (1991) assumed that the above timescale can be 
approximated by
the inverse of the growth rate of the unstable modes of the {\it
linearized} equations. By construction, the latter depends only on the 
source and  is {\it independent} of the 
{\it turbulence} it generates. The linear rate was used 
also by Canuto 
Goldman and Chasnov (1987,CGC) who, generalizing the work of Canuto and
Goldman (1985), proposed a model for fully developed turbulence.
 The linear rate
was also employed by Hartke Canuto \& Dannevik (1988) in the framework of
a DIA (Direct Interaction Approximation) model for turbulent convection
and by Canuto, Cheng, Hartke, \& Schilling (1991) for EDQNM (Eddy 
Damped Quasi-Normal Markovian) models.

The rate of energy input in the CM model must be improved since a fully
developed turbulence is expected to regulate the energy input from source
(buoyancy). The lack of feedback from the turbulence on the energy input,
prevents the model from being selfconsistent. However, because of the
complex structure of the EDQNM closure, the implementation of a
selfconsistent rate into the formalism of the CM model is not simple.
 Thus, we were forced  to simplify the structure of the
nonlinear interactions, so as to be able to formulate a workable
selfconsistent treatment. 

In  modeling the nonlinear interactions, we follow the
work of CGC (1987).  However, in the present model we generalize the
definition of the eddy correlation timescale, thus correcting some
shortcomings in the physics involved, and leading to an improved closure.
The resulting effective rate of energy input from the source (buoyancy)
depends now on both the {\it source} and the {\it turbulence}. As 
stated, the
present approach is complementary to that of the CM model. Here, the focus
is on the selfconsistent rate of energy input and not on the closure,
which is much simpler than that of the CM model.

In 
spite of the highly nonlinear structure of the model equations, it is 
possible to solve them {\it directly}  with no need for iterations. This 
reduced considerably the amount of numerical work required, and allowed
for an 
exploration of the model for a wide range of values of the convective 
efficiency, $S$, defined in equation (51). For each  
value of $S$ we obtained the spectral function which determines
 the turbulence bulk quantities. Thus, we
computed, as functions of $S$,
the convective flux, the turbulent viscosity, the turbulent mean 
squared velocity and the root mean squared turbulent 
pressure. 

We have applied the new model to the main sequence evolution of a 
solar model as well as to the evolutions of an extreme POP II chemical 
composition (Y=0.23, Z=$10^{-4}$) stars with $M\leq 0.9$ \Mo. The results 
are generally similar to those of the CM model. However, the 
new model has the
advantage that the overshoot required to fit the solar model is much 
smaller, in accord with recent empirical estimates,  and the 
ages of globular clusters are also somewhat smaller 
than the corresponding ages 
in the CM model (which in turn are smaller than  those derived within the 
MLT framework).

\section{The Model}
\subsection{The Rate Controlling Energy Input from the Source}

Let us consider a fully developed stationary turbulent convection 
characterized by the spectral functions $F(k)\,,G(k)$ and $H(k)$, of the 
turbulent velocity, temperature fluctuations and the turbulent 
convective flux, respectively. Before doing so, we write the dynamical 
time-dependent equations obeyed by these spectral functions (Yamaguchi 1963).

$$\parder ;t; F(k) +\nu k^2 F(k) = g\alpha H(k) +T_F(k) \ ,\eqno(1)$$

$$\parder ;t; G(k) +\chi k^2 G(k) = \beta H(k)+T_G (k) \ ,\eqno(2)$$
and

$$\parder ;t; H(k) +(\nu +\chi) 
k^2 H(k)= \beta \tau(k) F(k) + g\alpha \tau G(k) +T_H(k)\ ,\eqno(3)$$ 
where $T_F$, $T_G$, and $T_H$ denote the nonlinear transfer terms for the 
turbulent velocity, temperature, and convective flux. Here,  $g$ is the  
gravitational acceleration, $\alpha$ is 
the coefficient of thermal volume expansion at constant pressure 
(equaling $T^{-1}$ for an ideal gas), $\beta$ is the superadiabatic 
temperature gradient, $\nu$ is the microscopic viscosity, and $\chi$ is the 
microscopic thermometric conductivity appearing in the expression for the 
conductive flux
$$F_{cond} = -c_p \rho \chi {dT\over dz}\ .\eqno(4)$$
In stellar interiors, the dominant conductive flux is the radiative flux and
thus $\chi$ is the radiative thermometric conductivity.  The function $\tau(k)$ is 
given by

$$\tau(k)={x(k)\over 1+x(k)}\eqno(5)$$
with $x(k)=k^2_h/k^2_v$ measuring the anisotropy of the eddy 
corresponding to the
wavenumber $k$. Here, $k_h$ and $k_v$ stand for the horizontal and vertical 
wavenumbers, respectively (the vertical direction is that of the 
gravitational acceleration). 

In equation (1), $g\alpha H(k)$ plays the role of the 
energy source driving
the velocity fluctuations. More precisely, it equals the rate of energy 
per unit mass and per unit wavenumber, fed to the turbulence velocity 
field at wavenumber $k$. The  term $\nu k^2 
F(k)$ is the rate of energy per unit mass and unit wavenumber dissipated 
at  $k$ by the microscopic viscosity, while $-T_F(k)$ represent the rate 
of energy per unit mass and unit wavenumber transferred to
wavenumbers {\it other} than $k$. Analogous interpretations apply to 
equation (2) which 
describes the temperature fluctuations field.  

As stated above, we are interested in stationary turbulence. Thus, in 
equations (1)--(3) we set the time derivatives of the spectral functions 
equal to zero. Following CGC (1987), we assume that

$${T_H(k)\over H(k)}={T_F(k)\over F(k)}+{T_G(k)\over G(k)}\eqno(6)$$
which equations (1), (2), and (3) show to be equivalent to 
the assumption:

$$H(k)=\biggl(\tau(k) F(k) G(k)\biggr)^{1/2}\ .\eqno(7)$$
For equation (7) to 
be satisfied, the velocity and temperature fluctuations, at 
any  $k$, must be in phase. This is expected to be the case
for turbulent convection where the temperature plays the 
role of an {\it active scalar} whose fluctuations {\it drive} the 
velocity fluctuations.

We adopt the simplifying assumption that the transfer
terms in equations (1) and (2) describe transfer from small to large 
wavenumbers only (from large spatial scales to small ones).
While justified in three dimensional  turbulence where energy flows 
predominantly from large to small scales, this simplifying assumption 
neglects  non-local interactions (in $k$ space)
that give rise to reverse transfer (backscatter).

Denoting by $k_0$ the wavenumber 
corresponding to the largest eddy, we shall represent the transfer as

$$\int_{k_0}^k T_F(k') dk'=-\nu_t(k) y(k)\eqno (8)$$
where

$$y(k)=\int_{k_0}^k F(k') k'^2 dk'\eqno(9)$$
is the mean squared turbulent  vorticity and $\nu_t(k)$ is
the turbulent viscosity at wavenumber $k$, 
exerted by all eddies of smaller size (larger $k$). As such, it is 
expressed as an integral with limits $k$ and $\infty$. We shall return 
later
to the definition of the integrand.  Similarly,

$$\int_{k_0}^k T_G(k') dk'=-\chi_t(k) w(k)\eqno (10)$$
where 

$$w(k)=\int_{k_0}^k G(k') k'^2 dk'\eqno(11)$$
is the analog to the mean squared turbulent  vorticity
and $\chi_t(k)$ is
the turbulent conductivity at wavenumber $k$ 
exerted by all eddies of smaller size (larger $k$). With these closures, 
we obtain by integrating equations (1) and (2) 

$$g\alpha \int_{k_0}^k H(k') dk'=\biggl(\nu + \nu_t(k) \biggr) 
y(k)\ ,\eqno(12)$$ 
and

$$\beta\int_{k_0}^k H(k') dk'=\biggl(\chi + \chi_t(k) \biggr) w(k)\ 
.\eqno(13)$$ 
The last two equations allow us to  express the spectral function 
$G(k)$ in terms of $F(k)$. We obtain

$$G(k)= {\beta\over g\alpha}\lambda(k) F(k)\ ,\eqno(14)$$ 
with 

$$\lambda(k)={\biggl(y(k)\Sigma_t(k)\biggr)'\over y'(k)}\ ,\eqno(15)$$

$$ \Sigma_t(k)={\nu+\nu_t(k)\over \chi+\chi_t(k)}\  ,\eqno(16)$$
where a prime denotes a differentiation with respect to $k$.
Substituting  equation (14) in equation (7) yields

$$H(k)= \biggl({\beta\over g\alpha}\tau(k) \lambda(k)\biggr)^{1/2} 
F(k)\ .\eqno(17)$$ 
Thus, once $F(k)$ is known so are $G(k)$ and $H(k)$, implying 
that one needs to solve only for the spectral function $F(k)$. In 
order to do so, let us 
return to equation (12) and express $H(k)$ in terms
of $F(k)$ using equation (17). We obtain

 $$\int_{k_0}^k \biggl(n_s(k') +\nu k'^2\biggr) F(k') dk'=\biggl(\nu + 
\nu_t(k) \biggr) y(k)\eqno(18)$$
where $n_s(k)$ is the shorthand abbreviation 

$$n_s(k)\equiv -\nu k^2 + \biggl(g\alpha\beta 
\tau(k)\lambda(k)\biggr)^{1/2} \ , \eqno(19)$$ 
whose dimension are inverse time. 
The combination $n_s(k) +\nu k^2$ plays, in equation (18),  the role of the 
inverse of
the timescale  controlling the energy input into the turbulence at 
wavenumber $k$
from the source (buoyancy in the present case).  More specifically,  
 the {\it net rate of energy input} from the source per unit mass and unit 
wavenumber, at $k$, is $n_s(k) F(k)$. 

It is obvious, from equation (19), that $n_s(k)$ depends on the {\it 
source} (buoyancy) and on the {\it turbulent state}. Thus, it conforms 
with the requirements discussed in \S 1. In what follows we show that 
equations 
(18) and (19) allow us to  express $n_s(k)$ in terms of the turbulent 
viscosity $\nu_t(k)$. First, differentiate equation (18) with respect to 
$k$. The result is

$$n_s(k)-\nu_t(k)' {y(k)\over F(k)}=\nu_t(k) k^2\ . \eqno(20)$$
Next, use  equations (15), (20) and the definition of y(k), equation(9),
to get

$$\lambda(k)=\Sigma_t(k) +{\Sigma_t(k)'\over k^2 
\nu_t(k)'}\biggl(n_s(k)-\nu_t k^2\biggr)\ .\eqno(21)$$ 
With the help of equation (19), we transform equation (21) into a 
second order algebraic equation for $\lambda(k)$:

 $$\lambda(k)=\Sigma_t^2(k) {\chi'_t\over\nu'_t}+ \biggl(g\alpha\beta \tau(k) 
\lambda(k)\biggr)^{1/2} k^{-2} {\Sigma'_t\over\nu'_t}\ ,\eqno(22)$$ 
whose positive  solution is

 $$\lambda(k)={1\over 2}\biggl(g\alpha\beta \tau(k) \biggr)^{1/2} k^{-2} 
{\Sigma'_t\over\nu'_t} 
\left(1+\left(1+{4 k^4\over g \alpha\beta\tau}\left({\Sigma_t\over 
\Sigma_t'}\right)^2 \chi_t' \nu_t'\right)^{1/2}\right)\ .\eqno(23)$$ 
We adopt a relation between $\chi_t(k)$ and $\nu_t(k)$ 

$$\chi_t(k)= \bigg(\chi^2 + \sigma_t^{-2} 
\nu^2_t(k)\bigg)^{1/2} -\chi \eqno(24)$$ 
so that, as in CGC (1987), $\nu_t(k)/\chi_t(k)$ equals the constant 
$\sigma_t$ for 
large Pecle numbers, $Pe=\nu_t(k)/\chi$, while 
$\nu_t(k)/\chi_t(k)=2 \sigma_t^2 \chi/\nu_t(k)$, for small $Pe$.
 Using equation (24) to express $\Sigma_t$ and $\chi_t$ in terms 
of $\nu_t$, equation (23) now takes the form

$$\lambda^{1/2}(k)={1\over 2}(g\alpha\beta)^{1/2}\chi^{-1}\tau^{1/2}(k)k^{-2} 
{B(k)\over A(k)}\eqno(25)$$ 
with

$$A(k)=\bigg(1+\sigma_t^{-2} \nu_t^2(k)\chi^{-2}\bigg)^{3/2}\ ,\eqno(26)$$
and

$$B(k)=1-{\nu \nu_t(k)\over \sigma_t^2 \chi^2}+\left(\left(1-{\nu 
\nu_t(k)\over 
\sigma_t^2 \chi^2}\right)^2+A(k){4 
k^4\nu_t(k)\bigg(\nu+\nu_t(k)\bigg)^2\over 
g\alpha\beta\chi \sigma_t^2 \tau(k)}\right)^{1/2}\ .\eqno(27)$$
With this $\lambda(k)$, $n_s(k)$ of equation (19) becomes

 $$n_s(k)=-\nu k^2 +{g\alpha\beta\over 2\chi} {\tau(k)\over k^2} {B(k)\over 
A(k)}\ .\eqno(28)$$
Thus, we succeeded in expressing $n_s(k)$ in terms of the turbulent viscosity
$\nu_t(k)$, even though the latter {\it is still unspecified}.

\subsection{The Eddy-Correlation Timescale}

 In order 
to solve equation (18) (or equivalently equation (20)) we need an 
additional relation between the turbulent viscosity $\nu_t(k)$ 
and the spectral function $F(k)$. Without loss of generality, 
$\nu_t(k)$ can be written as 

$$\nu_t(k)=\int_k^{\infty} {F(k')\over n_c^*(k')} 
dk'\eqno(29)$$ 
where $n_c^*(k)$ has the dimensions of an inverse time. In 
order to determine $n_c^*(k)$ let us focus on its physical 
meaning. Differentiation 
of equation (29) yields that the turbulent viscosity contributed by 
eddies in the wavenumber interval $(k,k+dk)$ is

$$d\nu_t= n_c^*(k)^{-1} F(k) dk\ .\eqno(30)$$
Thus, $n_c^*(k)$ is proportional to the inverse of the eddy-correlation 
timescale at  wavenumber  $k$, or heuristically, the inverse of 
the  timescale for the eddy breakup. The eddy 
is damped because of two processes: interaction with the turbulent 
viscosity and the interaction with the source (microscopic viscosity is 
not considered here since in stars $\nu<<\nu_t$). 
 One can 
envisage the eddy as being "scattered" by "collisions" with the smaller 
eddies 
(turbulent viscosity) and by the source that drives energy into the eddy.
The  effective rate for the breakup (the inverse of the correlation 
timescale) will be taken to be the sum of the rates for the two processes 
as if they were operating independently: $\nu_t(k) k^2 + n_s(k)$, plus the 
sum of the rates for each process which is now affected by the other. 
Between "collisions" due to one process there is a 
random walk due to "collisions" of the other process. Thus, these
last two rates are 
$n_s(\nu_t k^2/n_s)^{1/2}$ and  $\nu_t k^2(n_s/(\nu_t k^2))^{1/2}$. 
Summing up the four rates  we obtain

$$\gamma n_c^*(k)= \nu_t(k) k^2 + n_s(k) + 2 \biggl(\nu_t(k) k^2
n_s(k)\biggr)^{1/2}=\biggl(\left(\nu_t(k) k^2\right)^{1/2} + 
n_s(k)^{1/2}\biggr)^2 \eqno(31)$$ 
where $\gamma$ is a proportionality constant, determined by the 
 normalization of $y(k)$ in the inertial range. In this range
equations (20), (29) and (31) yield

$$y(k)=\gamma^{-1} \left(\nu_t(k) k^2\right)^2\eqno(32)$$
while equation (18) results in

$$\epsilon=y(k) \nu_t(k) \eqno(33)$$
with $\epsilon=\epsilon(\infty)$ where

$$\epsilon(k)= \int_{k_0}^k \biggl(n_s(k') +\nu k'^2\biggr) F(k') dk'=\biggl(\nu + 
\nu_t(k) \biggr) y(k)\eqno(34)$$ 
is the energy rate per unit mass supplied to the 
turbulence from the driving source at all wavenumbers smaller than $k$.
Combined together, equations (32) and  (33) imply that

$$y(k)=\epsilon^{2/3}\gamma^{-1/3} k^{4/3}\ .\eqno(35)$$
This should coincide with $y(k)$ corresponding to
the Kolmogorov spectral function

$$y_{_K}(k)= {3\over 2} K_O \epsilon^{2/3} k^{4/3}\ .\eqno(36)$$
Thus, we obtain
$$\gamma=\left({2\over 3 K_O}\right)^3=0.0878 \left({K_O\over 
1.5}\right)^{-3}\eqno(37)$$
where $K_O$ is the Kolmogorov constant. We used here the same 
normalization with respect to $K_O$ as was used in CM (1991). 

With the definition of $n_c^*(k)$, equation (31),
 the model equations can be solved for $F(k)$, and consequently for $G(k)$
and $ H(k)$. We introduce for notational convenience (following the
convention of CGC (1987)) the rate $n_c(k)$ defined by

$$\nu_t k^2=\gamma n_c(k)\ . \eqno(38)$$
In the model for large scale turbulence  of Canuto \& Goldman (1985) $n_c^*$ 
was identified with $n_s$, neglecting the role of the turbulence. In CGC 
(1987) $n_c^*=n_c$ was used, neglecting the role of the source.
Here, $n_c^*$ depends both on $n_s$ and on $n_c$, as seen from equations 
(31) and (38), so it depends both on the source and on the turbulence. 
At $k_0$, $n_c^* = 4 n_c$  and therefore is $4$ times larger than in CGC 
(1987). For high
values of $k$ (practically few times $k_0$) $n_c^*\to n_c$. It is 
of interest to note that the inverse of the 
timescale for two-times velocity correlation according to the
DIA model, indeed conforms to the present definition of $n_c^*$
 (see figures 3 and 4 in Canuto \& Battaglia 1988). In particular at $k_0$ 
 it is indeed $\sim 4\ n_c$.

\subsection{Differential Equation,
Spectral Function, and Convective Flux}

From equations (29) and (38)  we have

$$F(k)=-\gamma n_c^*(k) \left(n_c\over k^2\right)'\ ,\eqno(39)$$
and equations (20) and (38) lead to

$$y(k)=n_c^*(k)\biggl(\gamma n_c(k) -n_s(k)\biggr)\ .\eqno(40)$$
Combining equations (39), (40) and (9) yields the differential equation for
$n_c(k)$

$$2 n_c^*(k) n_c'(k) + n_c^*(k)' \biggl(n_c(k)-\gamma^{-1} n_s(k)\biggr) 
-\gamma^{-1} n_c^*(k) n_s'(k)-2n_c(k) n_c^*(k){1\over k}=0 \eqno(41)$$
with $n_s(k)$ and $n_c^*(k)$ defined as functionals of $n_c(k)$ through
equations (28) and (31), respectively. A solution for $n_c(k)$ will
also yield these two rates,  as well as the spectral function $F(k)$ and 
the mean squared turbulent vorticity $y(k)$. 

The main objective of this work is the computation of the turbulent 
convective flux 

$$F_c=c_p\rho\overline {u_3 \theta}=c_p\rho \int_{k_0}^{\infty} H(k) dk=
c_p\rho\beta\chi\Phi\ . \eqno(42)$$
Here $u_3$ is the turbulent velocity in the flux direction, $\theta$ is 
the temperature fluctuation, and the overline denotes ensemble averaging. 
The dimensionless convective flux $\Phi$  is determined 
by equations (12) and (34) to be   

$$\Phi=\epsilon (g\alpha\beta\chi)^{-1}\ .\eqno(43)$$
The value of $\epsilon$ will be 
determined by using the fact that
$\epsilon(k)$ increases with $k$ and reaches its asymptotic value already 
in the inertial range. 
Thus, we shall follow the 
solution of $n_c(k)$ and the corresponding $\epsilon(k)$, from $k_0$ up 
to to the inertial range, until $\epsilon(k)$ saturates to $\epsilon$. 
From equations (34), (38) and (40) we 
find

$$\epsilon(k)=n_c^*(k) \left(\nu+{\gamma 
n_c(k)\over k^2}\right) \bigg(\gamma n_c(k)-n_s(k)\bigg)\ .\eqno(44)$$
thus, once $n_c(k)$ is obtained, $\epsilon(k)$ is determined too.

\section{Solution Procedure}

Turn now to the equation (41) which is a first order differential 
equation for $n_c(k)$ and thus a boundary value is required for a unique 
solution. Below we find the value of $n_c(k_0)$. The value of $k_0$ is 
determined by the width of the layer $\Lambda$, 

$$k_0={\pi\over \Lambda}(1+x_0)^{1/2}\eqno(45)$$
with $x_0\equiv x(k_0)$. Since by definition $F(k_0)=0$ and 
$y(k_0)=0$, equations (39), and (40) imply that

$$\gamma n_c(k_0)=n_s(k_0)=\nu_t(k_0) k_0^2\eqno(46)$$
and

$$\left({n_s\over k^2}\right)'_{k_0}=\left(\gamma n_c\over  
k^2\right)'_{k_0}=0\ .\eqno(47)$$ 
The last two equations together with equation (28) determine $n_c(k_0)$ 

$$\gamma n_c(k_0)=\sigma_t \chi k_0^2 z_0\ ,\eqno(48)$$
where $z_0$ is the solution of

$$\left({\sigma\over \sigma_t} +z_0\right)^2 
\left(1+z_0^2\right)={x_0\over1+x_0} \sigma_t^{-2} S_1^2\ .\eqno(49)$$
with

$$S_1=\pi^{-4} (1+x_0)^{-2} S=0.0045627 S\ ,\eqno(50)$$
where $S$ is the dimensionless product of the Rayleigh and Prandtl numbers

$$S= {g\alpha\beta \Lambda^4\over\chi^2}\ .\eqno(51)$$
$S$, or more precisely $S_1$, is a measure of the convective efficiency. 
With the determination of $n_c(k_0)$,
 equation (41) can be solved numerically to yield $n_c(k)$ for any 
value of $S$.

Since we deal with stellar interiors for which $\sigma=\nu \chi^{-
1}<<1$, we
take in all the equations the limit of $\nu\to 0$. Physically this means 
that the wavenumber at which the rate of dissipation by the microscopic 
viscosity becomes equal to $\gamma n_c(k)$, is much larger than $k_0$, so
that the spectrum exhibits 
an inertial range over many decades of $k$. 
For the limit of $\nu\to 0$, appropriate for stellar interiors, equation 
(49) yields an analytic solution for $z_0$ as function of $S$,

$$z_0={2^{1/2} \sigma_t^{-1}\tau(1) S_1\over 
     \sqrt{1 + \sqrt{1 + 4 S_1^2\tau^2(1)\sigma_t^{-2}}}}\eqno(53)$$
implying that

$$\gamma n_c(k_0)=n_s(k_0)=(g \alpha\beta)^{1/2} 
\eta_0(S)={\chi\over \Lambda^2} S^{1/2} \eta_0(S)\eqno(54)$$ 
with

$$\eta_0(S)={2^{1/2} \tau(1) S_1^{1/2}\over 
     \sqrt{1 + \sqrt{1 + 4 S_1^2\tau^2(1)\sigma_t^{-2}}}}\eqno(55)$$
and $\tau(1)=x_0(1+x_0)^{-1}$.
It is convenient to express the equations in terms of normalized rates

$$\eta_s(k)={n_s(k)\over n_s(k_0)}\ ,\eqno(56)$$

$$\eta_c(k)={\gamma n_c(k)\over  n_s(k_0)}\eqno(57)$$
and

$$\eta_c^*(k)={\gamma n_c^*(k)\over  n_s(k_0)}\eqno(58)$$
and to introduce a dimensionless wavenumber

$$q={k\over k_0}\ .\eqno(59)$$
Using equations (26), (27) and (28) in equation (47) results in

$$\left({\tau(k)\over k^4}\right)'_{k_0}=0\eqno(60)$$
 We adopt, following CGC (1987),

$$x(k)=\left(k \Lambda\over\pi\right)^2-1\eqno(61)$$
for which  equation (60) yields

$$k_0=\left(3\over2\right)^{1/2}{\pi\over\Lambda}\eqno(62)$$
Thus

$$\tau(q)=1-{2\over 3 q^2}\eqno(63)$$
and

$$x_0={1\over2}\eqno(64)$$
The differential equation for $\eta_c(q)$ is obtained from equation (41)

$$2 \eta_c^*(q) \eta_c'(q) + \eta_c^*(q)' (\eta_c(q)-\eta_s(q)) 
-\eta_c^*(q) \eta_s'(q)-2\eta_c(q) \eta_c^*(q){1\over q}=0 \ ,\eqno(65)$$
where now the prime denotes differentiation with respect to $q$. By 
definition [see eqns. (56), (57), (58)] 

$$\eta_c(q=1)=\eta_s(q=1)=1\ \ \ ;\ \eta_c^*(q=1)=4\ ,\eqno(66)$$
for any $S$. We have now

$$\eta_s(q)= {S_1^{1/2} \eta_0^{-1}\tau(q) B(q)\over 2 q^2 A(q)}\eqno(67)$$
where 

$$A(q)= \left(1 + {S_1 \eta_0^2 \eta_c^2(q))\over \sigma^2_t 
q^4}\right)^{3/2}\ ,\eqno(68)$$
and

$$B(q)=1 +\left(1+ {4 S_1^{1/2}\eta_0^3 A(q) 
\eta_c^3(q))\over \sigma^2_t \tau(q) q^2}\right)^{1/2}\ . \eqno(69)$$

The use of the normalized rates ensures that the computed quantities will
not be very small (large) even for very small (large) $S$ values, 
thus improving the numerical accuracy. Equation 
(44) becomes now,

$$\epsilon(q)= {1\over \gamma} (g\alpha\beta\chi) \eta_0^3 
S_1^{1/2}q^{-2}\eta_c(q)\eta_c^*(q) 
\biggl(\eta_c(q)-\eta_s(q)\biggr)\eqno(70)$$ 
the dimensionless convective flux is given by

$$\Phi(S)={1\over \gamma} \eta_0^3 
S_1^{1/2}q_f^{-2}\eta_c(q_f)\eta_c^*(q_f) 
\biggl(\eta_c(q_f)-\eta_s(q_f)\biggr)\eqno(71)$$
where $q_f$ is the upper value of q for which equation (65) is solved, 
and is well inside the inertial range (thus, $\epsilon(q_f)=\epsilon$).
 
\section{Results}
\subsection{Convective Flux}

We solved equation (65) for $\eta_c(q)$, with 
the initial 
condition of equation (66), for $S$ in the range $10^{-4}$---$10^{20}$ . 
We adopted the commonly used  value of $\sigma_t = 0.72$. Each
solution yields  $\eta_s(q)$ and $\eta_c^*(q)$ and the spectral function 
$F(q)$. For each such 
solution we computed $\epsilon(q)$, followed it to saturation, and thus
obtained $\epsilon(S)$ and $\Phi(S)$.

In Figure 1 we show $\epsilon(q)$ in units of $(K_O/1.5)^3 g \alpha \beta 
\chi$, for $S=10^6$. The qualitative behavior of $\epsilon(q)$ is typical 
for any 
value of $S$: it starts from zero and saturates in the inertial range to 
$\epsilon$. From equations (70)  and (71) it follows that the asymptotic 
value of the graph equals $\Phi$ in units of
$(K_O/1.5)^3$. 

In Table 1  we list  $\Phi(S)$ (rounded to 4
figures) for $20$ representative values of $S$. 
From 
equation (71) it follows that $\Phi(S)\propto \gamma^{-1}\propto K_O^3$. The 
values shown in Table 1 are in units of $\left(K_O/1.5\right)^3$.
The limiting behavior of $\Phi(S)$ is given by

$$\Phi= 2.65 \times 10^{-5}\left({K_O\over 1.5}\right)^3  S^2\ \ \ \ ; 
\ \ S<<1\ , \eqno(72)$$ 
and

$$\Phi=1.6853 \left({K_O\over 1.5}\right)^3 
\left({\sigma_t\over 0.72}\right)^{3/2} S^{1/2}\ \ \ \ ; \ \ S>>1 
\ .\eqno(73)$$

\begin{deluxetable}{rrrrrrr}
\small
\tablecaption{The Results of the Model}
\tablehead{
\colhead{$S$} & \colhead{$\Phi$} & \colhead{${\Phi\over \Phi_{MLT}}$} & 
\colhead{$\Phi\over \Phi_{CM}$} &\colhead{$\overline {v^2}$} & 
\colhead{$p_t$} & \colhead{$C$}}

\startdata
$10^{-4}$&$2.65\times 10^{-13}$&$0.3091$&$2.814$ &$2.112\times 10^{-11}$&
$5.3408\times 10^{-12}$&$0.2529$\nl
$ 0.01$&$2.65\times 10^{-9}$&$0.3092$&$2.814$ &$2.112\times 10^{-7}$&
$5.341\times 10^{-8}$&$0.2529$\nl
$1.$&$2.65\times{{10}^{-5}}$&$0.3148$&$2.819$ & $2.112\times 10^{-3}$ & 
$5.341 \times 10^{-4}$ &$0.2529$\nl 
$10$&$2.649\times 10^{-3}$&$0.3663$&$2.863$ & $0.2111$ & 
$0.5339$ &$0.2529$\nl
$30$&$2.38\times 10^{-2}$&$0.4811$&$2.96$ & $1.895$& $0.4793$& $0.2529$\nl
$ 10^2$&$0.2557$&$0.8745$&$3.203$ & $20.52$&$5.186$ &$0.2527$\nl
$300$&$1.88$&$1.806$&$3.409$ & $157.6$ & $39.63$ & $0.2515$\nl
$10^3$&$10.23$&$3.334$&$3.045$ & $1.019\times 10^3$ & $252.4$ & $0.2478$ \nl
$ 10^4$&$85.01$&$5.82$&$2.065$ & $1.685\times 10^4$ & $4.017 \times 10^3$ 
&$0.2384$\nl 
$10^5$&$388.5$&$7.381$&$1.59$ & $1.954 \times 10^5$ &$4.515\times 10^4$ & 
$0.2311$\nl
$10^6$&$1.442\times 10^3$&$8.312$&$1.348$ & $2.049 \times 10^6$ & 
$4.648 \times 10^5$&$0.2268$\nl 
$10^7$&$4.917 \times 10^3$&$8.849$&$1.209$ & $2.085 \times 10^7$ &
$4.680 \times 10^6$ & $0.2245$\nl 
$10^8$&$1.615\times 10^4$&$9.155$&$1.125$ & $2.094\times 10^8$ & 
$4.687 \times 10^7$ &$0.2238$\nl
$10^9$&$5.21 \times 10^4$&$9.326$&$1.072$ & $2.099\times 10^9$ & $4.688 
\times 10^8$ & $0.2233$\nl 
$10^{10}$&$1.665\times 10^5$&$9.42$&$1.038$ & $2.101\times 10^{10}$ 
& $4.688 \times 10^9$ &$0.2231$\nl
$10^{12}$&$1.679\times 10^6$&$9.498$&$1.001$ & $2.101 \times 10^{12}$ & 
$4.688 \times 10^{11}$ & $0.2231$\nl
$10^{14}$&$1.684\times 10^7$&$9.528$&$0.9863$ & $2.101 \times 10^{14}$ & 
$4.689 \times 10^{13}$ & $0.2231$\nl
$10^{16}$&$1.685\times 10^8$&$9.534$&$0.9795$ & $2.101 \times 10^{16}$ & 
$4.689 \times 10^{15}$ & $0.2231$\nl
$10^{18}$&$1.685\times 10^9$&$9.534$&$0.9764$ & $2.101 \times 10^{18}$ & 
$4.689 \times 10^{17}$ & $0.2231$\nl
$10^{20}$&$1.685\times 10^{10}$&$9.534$&$0.9751$ & $2.101 \times 10^{20}$ & 
$4.689 \times 10^{19}$ & $0.2231$\nl

\enddata

\tablenotetext{}{$\Phi$ is in units of $(K_O/1.5)^3$ and its high $S$
limit scales as $(\sigma_t/0.72)^{3/2}$. $\overline {v^2}$ and $p_t$ are in 
units of $\left(\chi\over
\Lambda\right)^2 \left({K_O\over 1.5}\right)^3$ and $\rho \left(\chi\over
\Lambda\right)^2 \left({K_O\over 1.5}\right)^3$, respectively. Their high
$S$ limits scale as $(\sigma_t/0.72)$. The dimensionless ratio, C, is 
defined in equation (88).} 
\end{deluxetable}

In applications of $\Phi(S)$ to stellar structure codes, it is useful to 
have an analytic fit formula to the convective flux. We derived such a 
fit with a deviation $\les 3\%$:

$$\Phi=F_1(S)\, F_2(S)\eqno(74)$$ 
where

$$F_{1}(S)=\left({K_O\over 1.5}\right)^3 a \, 
S^{k}\biggl(\left( 1 + b\, S \right)^{m} - 1
        \biggr)^{n}\eqno(75) $$
where
$$a=10.8654\ ,b=0.00489073\ ,k=0.149888 \ ,m=0.189238 \ ,n=1.85011$$
and

$$F_{2}(S)=1 + {c\, S^{^p}\over {1 + d\, S^{^q}}} + 
  {e\, S^{^r}\over {1 + f\,  S^{^t}}}\ ,\eqno(76)$$
where
$$c=0.0108071 \ ,d=0.00301208 \ , e=0.000334441 \ , f=0.000125\ ,$$

$$p=0.72 \ , q=0.92 \ , r=1.2 \ , t=1.5 \ .$$

In order to judge the quality of the fit, in Figure 2  we display the 
ratio between the fit function and an interpolation of the numerical values, 
as a function of log $S$.

\
Also shown in Table 1 are the ratios $\Phi(S)/\Phi_{MLT}(S)$ and 
$\Phi(S)/\Phi_{CM}(S)$
where $\Phi_{MLT}$ and $\Phi_{CM}$ are the values corresponding to the 
MLT,

$$\Phi_{MLT}(S)={729 \over 16} S^{-1}\left(\left(1+{2\over 81} S\right)^{1/2}
-1\right)^3\ ,\eqno(77)$$
and to the CM model (their eq. [32]).
 Figures 3 and 4
display $\Phi(S)/\Phi_{MLT}(S)$ and $\Phi(S)/\Phi_{CM}(S)$, respectively. 
Note
that the qualitative behavior of $\Phi(S)/\Phi_{MLT}(S)$ is similar to 
that of the CM model---higher flux for high $S$ and lower flux for low $S$ 
values. The comparison with $\Phi_{CM}(S)$ shows that while the two models
yield essentially the same flux for high $S$ values, the new model 
predicts higher fluxes for intermediate and low $S$ values, and the flux 
ratio is maximal for $S\sim 300$. Comparisons of 
$\Phi(S)$ computed within the new model with $\Phi(S)$ computed in a model 
with the  same definition of $n_c^*$ but with $n_s$ equal to the linear 
growth rate, indicates that the above
local maximum at $S\sim 300$ is a feature resulting from the use of the  
selfconsistent rate. 
We recall that $S_1=0.0045627\, S$ rather than $S$ is the measure of the 
convective efficiencies. Thus, the borderline between low and high 
efficiencies is around $S=300$, which is also where the ratio of the new 
convective flux to that of the CM model is maximal.

\subsection{Turbulent Velocity, Turbulent Pressure and Turbulent Viscosity}
The mean squared turbulent velocity is defined by

$$\overline {v^2}=\int^{\infty}_{k_0} F(k) dk\eqno(78)$$ 
which with the use of equation (39) can be expressed as

$$\overline {v^2}=\chi^2 \Lambda^{-2}\gamma^{-1} {1\over \pi^2 (1+x_0)} S 
\eta_0^2(S) \int^{\infty}_1 f(q) dq\ ,\eqno(79)$$ 
with

$$f(q)=-\eta_c^*(q) \left({\eta_c(q)\over q^2}\right)'\ .\eqno(80)$$ 
We computed $\overline {v^2}$ for each value of $S$ and the results 
 are presented in Table 1. The limiting behavior is given by

$$\overline {v^2}= 2.1117\times 10^{-3} \left({K_O\over 1.5}\right)^3 
\left({\chi\over\Lambda}\right)^2 S^2\ \ \ ; \ \ \ S<<1 \eqno(81)$$
and

$$\overline {v^2}= 2.10146 \left({K_O\over 1.5}\right)^3  
\left({\sigma_t\over 0.72}\right) 
\left({\chi\over\Lambda}\right)^2 S\ \ \ ; \ \ \ S>>1 \eqno(82)$$

The turbulent 
pressure is of importance in helioseismological models. 
Batchelor (1953) derived an expression for the mean squared turbulent 
pressure (for the case of isotropic turbulence) in terms of the spectral 
function $F(k)$

$$p_t^2= {1\over 4} \rho^2 \int^{\infty}_{k_0} \int^{\infty}_{k_0} 
F(k) F(k') I(k/k') dk'dk\eqno(83)$$
where $\rho$ is the mean density and the dimensionless integral $I(x)$
is 

$$I(x)=I(1/x)={1\over 2}\left(x^2+x^{-2}\right) -{1\over 3} -{1\over 4} 
\left(x+x^{-1}\right)\left(x-x^{-1}\right)^2 {\rm ln}{1+x\over 
|1-x|}\ .\eqno(84)$$
$I(x)\to 0$ for $x\to 0\,,\infty$ and is maximal at $x=1$, where it 
equals $2/3$. Thus, the pressure is mostly contributed by $k'\sim k$ 
which are close to the maximum of $F(k)$. Using equation (39) we obtain

$$p_t^2= {1\over 4\gamma^2} \rho^2 \chi^4 \Lambda^{-4} {1\over 
\pi^4 (1+x_0)^2} S^2 \eta_0^4(S) \int^{\infty}_1 \int^{\infty}_1 
f(q) f(q') I(k/k') dq'dq\ .\eqno(85)$$

The computed values of the root mean squared turbulent pressure
are displayed in Table 1. The asymptotic behavior is given by

$$p_t= 5.3408 \times 10^{-4}
\rho \left(\chi\over \Lambda\right)^2 \left({K_O\over 1.5}\right)^3  S^2\ 
\ \ \ ; \ \ \ S<<1\eqno(86)$$
and

$$p_t= 0.4689
\rho \left(\chi\over \Lambda\right)^2 \left({K_O\over 1.5}\right)^3  
\left({\sigma_t\over 0.72}\right)  S\ \ \ \ ; \ \ \ S>>1\ .\eqno(87)$$

A quantity of interest is $C(S)$ defined as

$$C(S)={p_t\over \rho \overline 
{v^2}} \eqno(88)$$

The computed values of $C$ are listed in Table 1. As can be seen, it is
almost constant for all values of $S$, ranging from $0.253$ for low $S$ to
$0.223$ for high values of $S$. 

As with the convective flux it is useful to have analytical fit formulae
for the mean squared turbulent velocity and for the root mean squared
turbulent pressure. We derived such fits which represent the numerical 
values with
precision better than $3\%$. For the mean squared velocity we derive the
fit

$$v^2=\left({\chi\over\Lambda}\right)^2 F_3(S)\, F_4(S)\ ,\eqno(89)$$
where

$$F_3(S)=\left({K_O\over 1.5}\right)^3 {0.00101392\,S^2\over {1 + \sqrt{1
+ 0.000017848\,S^2}}} \eqno(90)$$ 
and

$$F_4(S)=
6.39899 + {2.256815\,
      \left( -1. + 0.000777055\, S^{0.868589} \right)
     \over {1. + 0.000777055\, S^{0.868589}}} \ .\eqno(91)$$
Similarly, for the root mean square of the turbulent pressure we find

$$p_t= \rho \left(\chi\over \Lambda\right)^2 F_3(S)\, F_5(S)\eqno(92)$$
with $F_3(S)$ given by equation (90) and

$$F_5(S)=1.49168 + 0.45185{ -1. + 0.00111378 \, S^{0.868589} \over   
1. + 0.00111378 \, S^{0.868589}}\ . \eqno(93)$$

Finally, the turbulent viscosity, is given in the present model already
in an {\it analytic} form. From equations (38), (44), (45), (54) and (55) we 
find that
 
$$\nu_t\equiv \nu_t(k_0)= \chi {0.00215086\,S\over
   \sqrt{1 + \sqrt{1 + 0.000017848\,S^2}}}\ .\eqno (94)$$
The limiting behavior
 of  $\nu_t$ is given by

$$\nu_t=  0.00152089\, \chi\, S \ \ \ \ ; \ \ S<<1 \eqno(95)$$
and 

$$\nu_t=  0.03309\, \chi\, S^{1/2}\ \ \ \ ; \ \ S>>1. \eqno(96)$$

\section{Application to stellar models}

The new convective fluxes have been included in the ATON stellar
structure code (for an update on the physical and numerical details of the
code, see Mazzitelli et al. 1995 and references therein). We have 
computed  the main 
sequence evolution of a solar model as well as a set of
evolutions for Pop II stars having $M \le 0.9 M_{\odot}$,
from the zero age main sequence to the base of the red giant phase.

Before turning to a detailed discussion of the results, we recall that,
the turbulent length-scale 
 $\Lambda$ at a given depth $z$ inside a convective region, must
 also include the thickness $OV$ of the overshooting layers (if any) beyond
the formal Schwarzschild boundary (see D'Antona and Mazzitelli 1994).
At present,  the $OV$ phenomenon has not yet been 
 fully quantified in a reliable way (Umezu 1995) even though the underlying 
equations have been derived (Canuto 1993). However,
empirical evidence 
from comparisons between stellar models constructed with local convection
theories, and observations of intermediate mass main sequence stars in young
open clusters, suggests quite stringent limits on the extent 
of the $OV$, namely,  $0 \le OV \le
0.2 H_p$ (Stothers \& Chin 1992). Lacking a formal theory, we shall write 

$$\Lambda = z + \alpha^* H_p^{top}\ .\eqno(97)$$
where $H_p^{top}$ is the pressure scale height at the upper
 boundary of the convective layer determined by the Schwarzschild 
criterion and $\alpha^*$, which should not exceed
 $0.2$, can be regarded as a {\it fine-tuning} parameter. 

We stress that the role
 of $\alpha^*$ in the CM and in the present models  is {\it radically} 
different from the role of the parameter $\alpha$ in the MLT model where 
$\Lambda=\alpha H_p(z)$.
In fact, $\alpha$ is a free adjustable parameter through which
modelists try to capture all the physical uncertainties (e. g. 
opacities, convection, thermodynamics etc.) relevant to the 
evaluation of the effective temperatures of stars . Unfortunately, not 
only does this procedure seriously hinder  the predictive power of
stellar modeling
($\alpha$ is determined {\it a posteriori}), but the fit to the observed
surface temperature does not automatically guarantee that also the {\it 
internal temperature profile} is correct.
In this context, it is worth noting that Baturin \& Mironova (1995) and  
 Monteiro et al. (1995)
have recently shown that the solar internal temperature profiles 
predicted by the CM 
model are in better agreement with helio-seismological data than those 
derived from 
the MLT. Moreover, Gabriel (1995) has shown that  the
MLT could be made to predict a CM-like internal temperature profile, 
provided that $\alpha$ is {\it forced to
vary inside the convective region}, in a manner that represents an a posteriori fitting. This 
quite clearly  shows that a) the MLT has no predictive power and 
b) the degree of artificiality that is required to make the MLT yield 
results that the CM model produces quite naturally. 
The parameter $\alpha^*$, on the other hand, quantifies a well identified
{\it physical process}, the convective overshooting $OV$ and, as seen in
the following, in the present model, only a marginal amount of tuning is
allowed anyhow. 

Finally, the convective flux in Table 1 is normalized to a 
value of the
Kolmogorov constant of $K_O=1.5$, as in the CM model. Since  recent
experimental data suggest higher
values of $K_O$ (up to $\sim 1.9$), we employed for the stellar modeling a 
fiducial value of $K_O=1.7$  in the fit-formula, equation (75). Since, as 
discussed in \S 4.1, the convective 
flux scales as $K_O^3$, the flux used is a factor of $(1.7/1.5)^3$ larger 
than the numerical values in Table 1. In Mazzitelli et 
al. (1995) $K_O=1.8$ was used.

Using the ATON code, and updating the low--T opacities
according to Alexander and Ferguson (1994), we obtained a fit to 
the observed solar
radius and luminosity at an age $\sim 4.55$ Gyr (Bahcall 1989), and with a
metal abundance Z=0.0175 (Grevesse and Noels 1993), with
Y$\sim$0.27 and $\alpha^* \sim 0.08$. The latter corresponds to a very
small amount of overshooting of a {\it few kilometers}, and is in full 
accordance with the observational limits of Stothers \& Chin (1992). On the 
other hand, had  we
employed the original CM model the required value would have been 
$\alpha^*\sim 0.2$ which is a borderline value.
To find the maximal variance in  $T_{eff}$ allowed by the $\alpha^*$  
parameter,
we computed two solar models with the borderline values  
$\alpha^*=0$ and $\alpha^*=0.2$. The difference in $T_{eff}$
between these last two models turned out to be $<$4\%.

Figure 5 shows the internal profiles of the dimensionless 
temperature
gradient, $d\,{\rm log}\,T/d\,{\rm log}\,P$, in the region of the 
overadiabaticity peak for solar models 
computed with the present, the CM and MLT models (for the MLT model 
$\alpha =1.55$). The
results of the present model are very similar to those of the CM model but
 are quite different from the MLT results.
The similarity of the results of the present and CM models, in spite of the 
 difference (factor $\sim 3$) in the value of the convective flux 
for intermediate and low convective efficiencies, can be readily 
understood. In
the more external convective layers, the density is 
quite low ($\rho \le 10^{-7}$), and ${\rm Log}\,S<2$. The turbulent 
flux is also
quite low, and convective energy transfer is very inefficient. Most of
the flux in this region is therefore carried by radiation, and the 
temperature gradient sticks to the radiative one.
In deeper convective layers, where $\rho$, $S$ and  $\Lambda$ are larger,
convection begins to be efficient and the value of the
temperature gradient is determined by the turbulent convective flux. 
However, for $S$ values such that Log $S \gg 3$, the new fluxes are very 
close to  the CM ones, and so is the resulting temperature
gradient. The vicinity of the gradient peak, 
 ${\rm Log}\, S\sim 2$, is close to the $S$ value where the 
present and CM models differ the most ( see Fig. 3). Thus, this is the 
region where we can expect some sizable difference between  the temperature 
profiles, as Figure 5 indeed shows.

Because of 
the similarity of the results from the present and the CM models, the 
experimental benchmark provided by
helioseismological data cannot discriminate between the two models
(Antia and Basu 1995). However, on the basis of  stellar modeling, 
 we stress
that the new fluxes require a lower value of $\alpha^*$ to fit the sun, 
which is more in agreement with the results of Stothers \& Chin. Since the 
low--T radiative opacities are probably still slightly
underestimated, and larger opacities require a larger $\alpha^*$ to 
fit  the
sun, we  prefer the present new fluxes over the original CM values. The 
reason is that the 
latter, once the updated values of low--T opacities become available, could 
require values of
$\alpha^*$ larger than allowed by the observational upper limit on 
overshooting.

As a further check, we have also applied the new fluxes to the computation of
evolutionary tracks and isochrones for stars with $Y=0.23$ and $Z=10^{-4}$.
The isochrones are shown in Figure 6, together with the fiducial sequence of
the globular cluster M68. Details on the computations of both tracks and
isochrones, on the observational to theoretical correlations, as well as 
the chemistry, reddening and distance modulus for M68 can be found in 
Mazzitelli
et al. (1995). Here we simply recall that the apparent ``kinks'' in the
isochrones are a true physical feature, which is expected and explained 
within the CM framework, and which exists also in the present model.

The age of the cluster is in the range $11 \div 12$ Gyr, somewhat younger 
than the $12
\div 13$ Gyr found with the CM fluxes,  which itself is younger than the $13 
\div 15$
Gyr derived within the MLT. Whether this difference is 
significant for solving the age  conflict of
globular clusters with the age of the universe, following
from the recent determinations of high values of $H_0$ (Freedman et al.
1994, Pierce et al. 1994), is beyond the purposes of the present paper. 
This point will be discussed elsewhere (Canuto et al. 1996). 

In conclusion, the application of the convective fluxes of the present model
to stellar structure does not alter appreciably the main results of the 
CM model. Nevertheless,
the new model is  preferable on two grounds. From the 
theoretical 
viewpoint, it determines the rate of input of energy from 
buoyancy to the turbulence in a more physically consistent manner. From 
the astrophysical point of view, it requires a smaller extent of 
overshooting, which is in better agreement with recent observational 
results.

\section{Discussion}

We have presented a selfconsistent model for turbulent convection
 based on a simplified treatment of the non-linear interactions among the
eddies. The important novel feature of the present model is the
formulation of a selfconsistent rate for energy input from the {\it
source} (buoyancy) into the turbulence, which depends both on the source
parameters and on the {\it turbulence} itself. This represents an
improvement compared to the CM model where the rate of energy input was
the growth rate of the {\it linear} unstable modes. The focus of the
present model is on the selfconsistent rate of energy input at the expense
of a less complete treatment of the non-linear eddy interactions. The
latter is much simpler than in the CM model and describes  transfer only from
small to large wavenumbers. This representation
 neglects  non-local (in $k$ space) interactions  between 
the eddies that lead to a reverse transfer (backscatter) that are 
included in the CM model. 

 We have explored the model for a wide range of convective efficiencies and
computed, numerically, the dimensionless convective flux, the turbulent
squared velocity and the root mean squared turbulent pressure, as 
functions of the convective efficiency
$S$. The results were fitted by analytical formulae with precision better
than $3\%$ over the range $S=10^{-4}\div 10^{20}$. The turbulent viscosity
in the model is already given by an analytic expression. 
The convective flux, is larger than that of the MLT for high convective 
efficiencies and lower than it for low convective efficiencies. This general
behavior is similar to that of the CM model. The high $S$ fluxes are 
very close to those predicted by the CM model but the intermediate and 
lower $S$ fluxes are larger than those of the CM model.

It is of interest to note that even the very simple CGC (1987) model
shares the same qualitative behaviour of the convective flux relative to
that of the MLT flux, as function of the convective efficiency. The fact
that three models differing in their treatment of the energy input rate
and in the modeling of the nonlinear transfer, still yield
 the same  behaviour is quite intriguing. It suggests that  accounting 
for the full turbulence spectrum (common to all three models) is 
far more important than the detailed way in which thye latter is done.

 We have applied the new model to the main sequence evolution of a solar 
model as well as to evolutions of POP II stars with $M\leq 0.9$ 
\Mo. The convective turbulent length scale was taken equal to 
the depth in the convective zone, as in the CM model.
The results are generally similar to those of the CM model. However, the new 
model has the
advantage that the overshooting required to fit the solar model is 
smaller than in the CM model. Also, the 
ages of globular clusters are smaller than the corresponding ages 
in the CM model by $\sim 1$ Gyr, which  may help alleviate a
possible conflict between the ages of globular clusters and a high value of 
$H_0$.
As already noted in \S 5, the similarity between the temperature profiles 
 for the solar model, predicted by the present and 
the CM models, renders the models practically indistinguishable by 
helioseismological data (Antia \& 
Basu 1995). The situation is similar with regard to solar atmosphere 
modeling (Kupka 1995).  However, atmospheres of cool stars are expected to
yield observable differences between the two models (Kupka 1995).

From the theoretical perspective, a more complete model is one that 
incorporates a selfconsistent rate of energy input
while  keeping the non-linear interactions in their full generality, as done
within the CM model. Work in this direction is in progress.

Finally, the present and the CM models are based on two-point correlations
of the turbulent quantities. This methodology, preferred by the physics
community, yields information about the spectral properties of the
turbulence. Yet, its applicability to inhomogeneous and anisotropic cases
is limited. An alternative approach, based on one-point correlation
functions, is widely used in the engineering community and can handle
anisotropy and inhomogeneity. While the spectral information is lost in
this Reynolds Stress formalism, the method is easy to apply to non-local
and space-dependent problems and has the potential to treat
 stellar  convective overshooting (Canuto  1993). The method has already
 been successfully applied to the planetary boundary layer (PBL) which is
the seat of strong convection (Canuto  et al. 1994a) as well as to study
the interaction between shear, vorticity and buoyancy at the surface of
the sun (Canuto et al. 1994b).  Work is in progress (Gabriel 1996, Houdeck
1996) to apply the same method to the study of helioseismology.

\bigskip

\acknowledgments

This work was supported by the US-Israel BSF grant 94-00314, to I. Goldman
 


\begin{figure}
\plotone{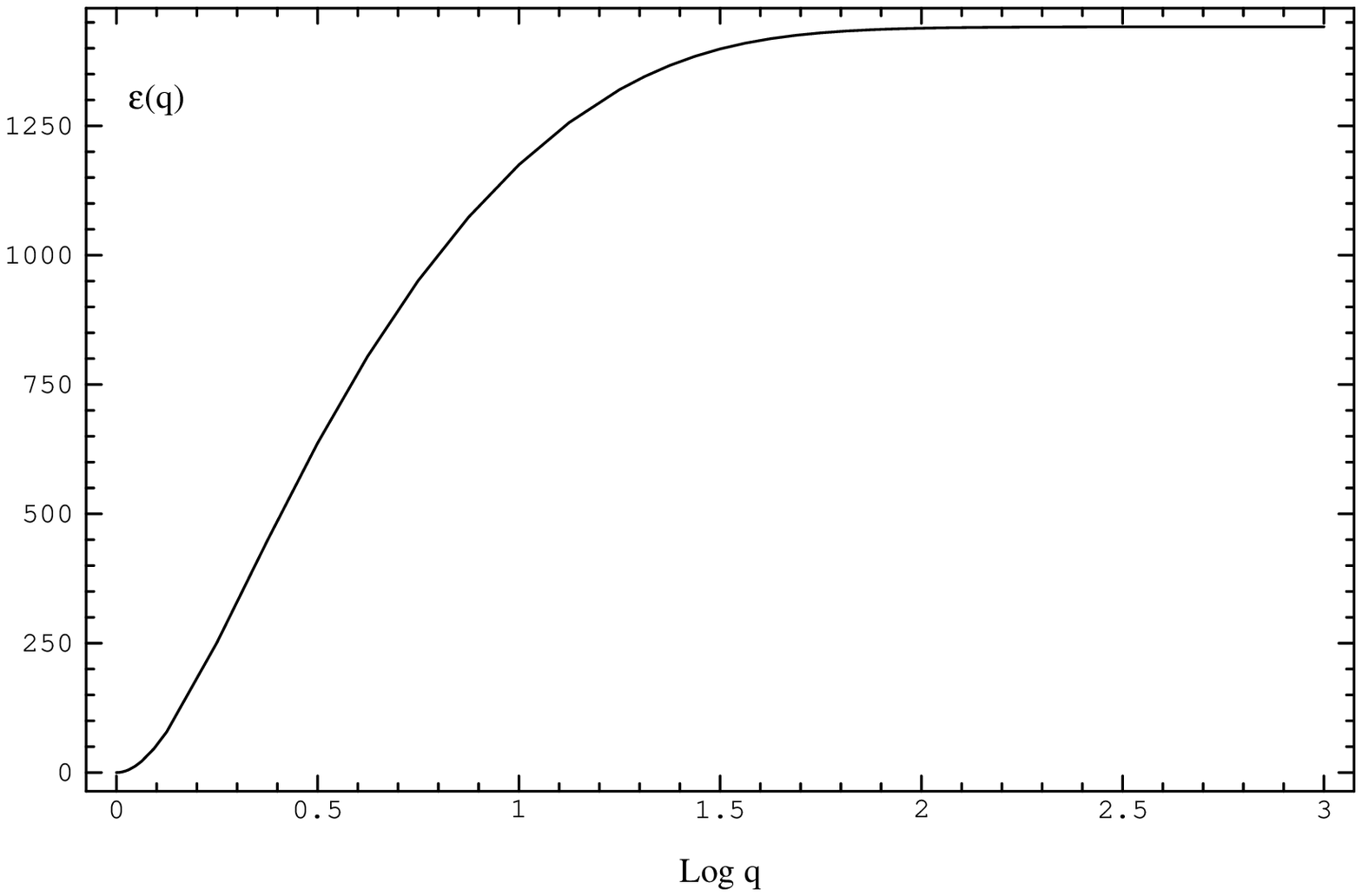}
\vskip - 5 truecm
\caption{$\epsilon(q)$ in units of $(K_O/1.5)^3 g \alpha \beta \chi$,
for $S=10^6$. The asymptotic value shown is actually $\Phi$ in units of
$(K_O/1.5)^3$.}
\end{figure}

\begin{figure}
\plotone{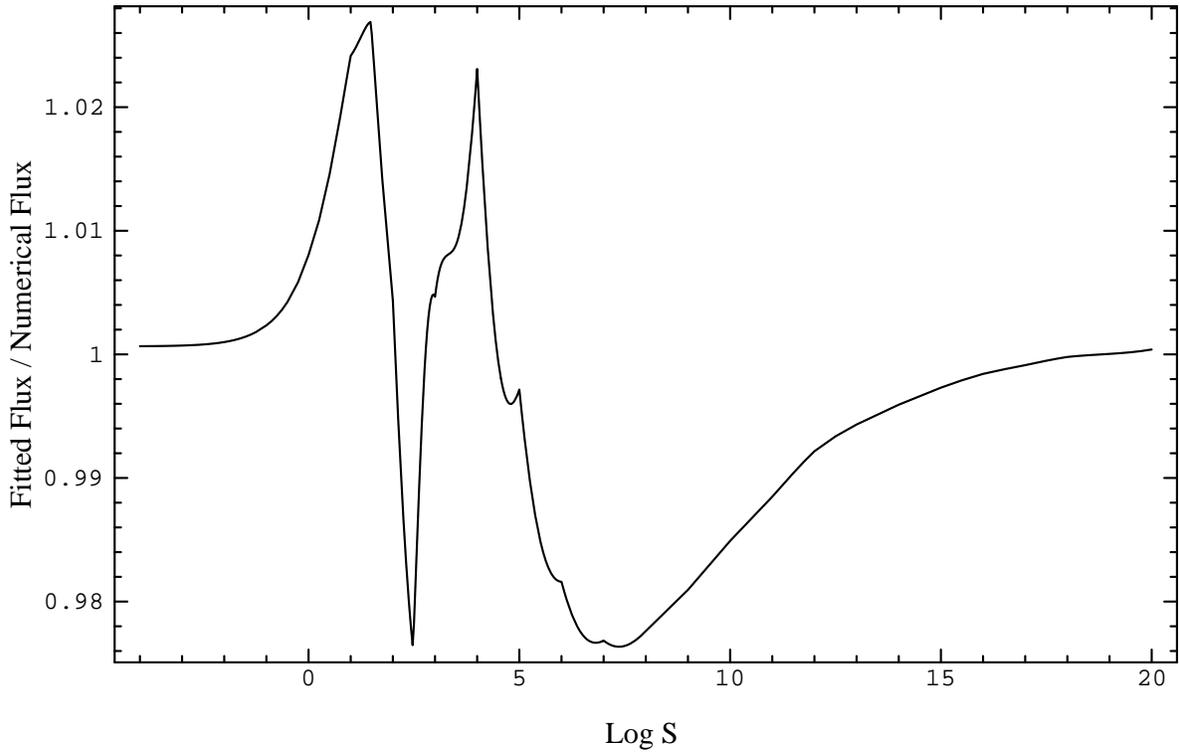}
\vskip - 5 truecm
\caption{The ratio between the fit function for $\Phi(S)$, eq. (74), and
 an interpolation of the numerical values of
$\Phi(S)$.}
\end{figure}   

\begin{figure}
\plotone{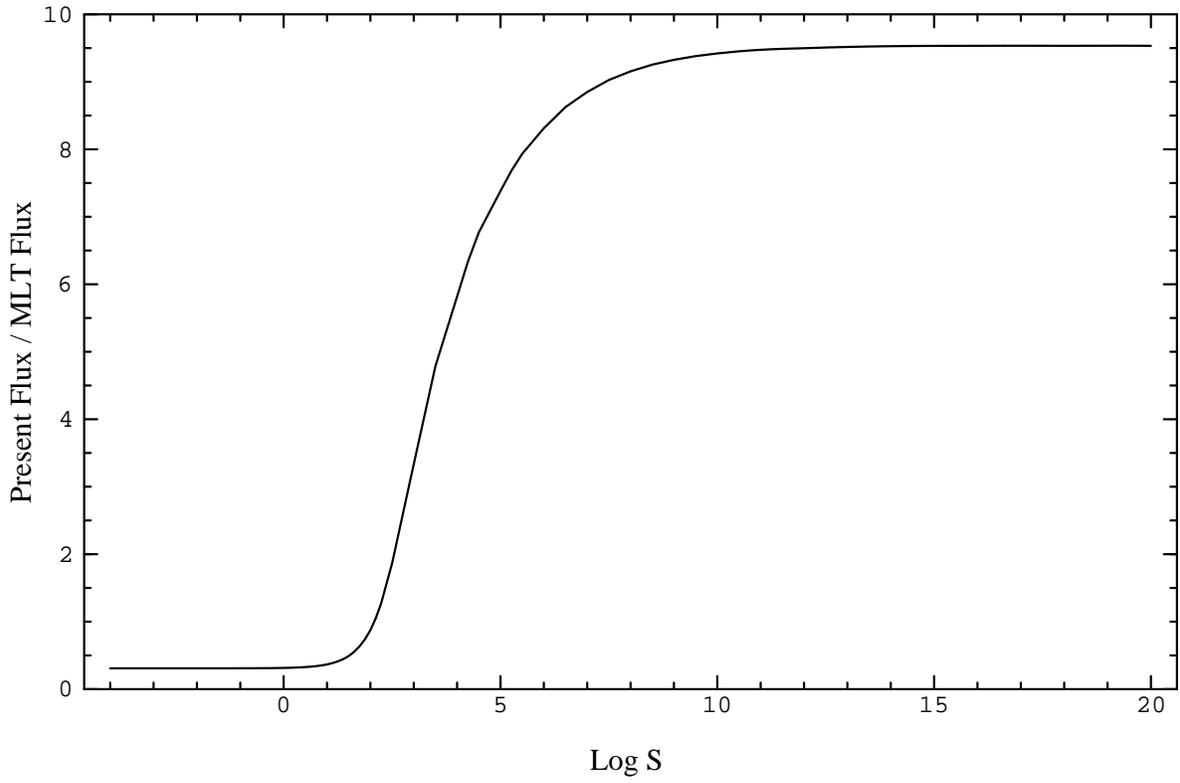}
\vskip - 5 truecm
\caption{The ratio $\Phi(S)/\Phi_{MLT}(S)$.}
\end{figure}

\begin{figure}
\plotone{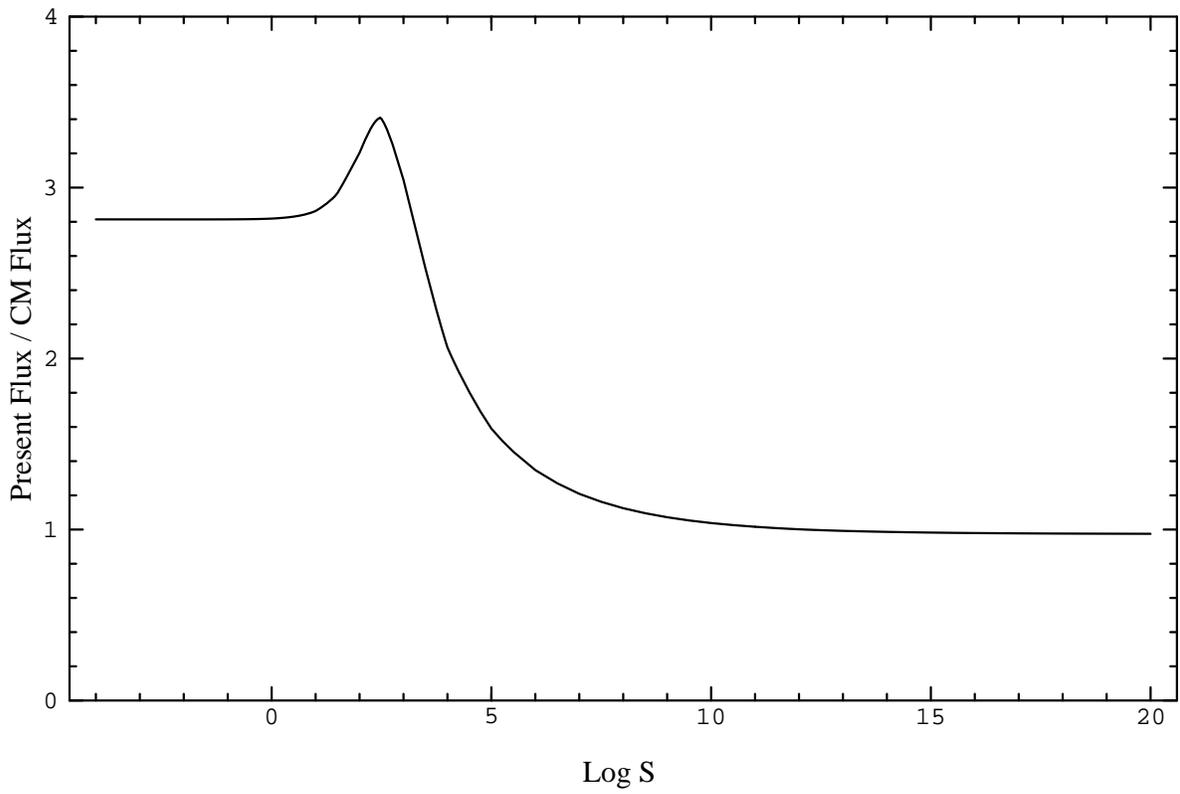}
\vskip - 5 truecm
\caption{The ratio $\Phi(S)/\Phi_{CM}(S)$.}
\end{figure}

\begin{figure}
\plotone{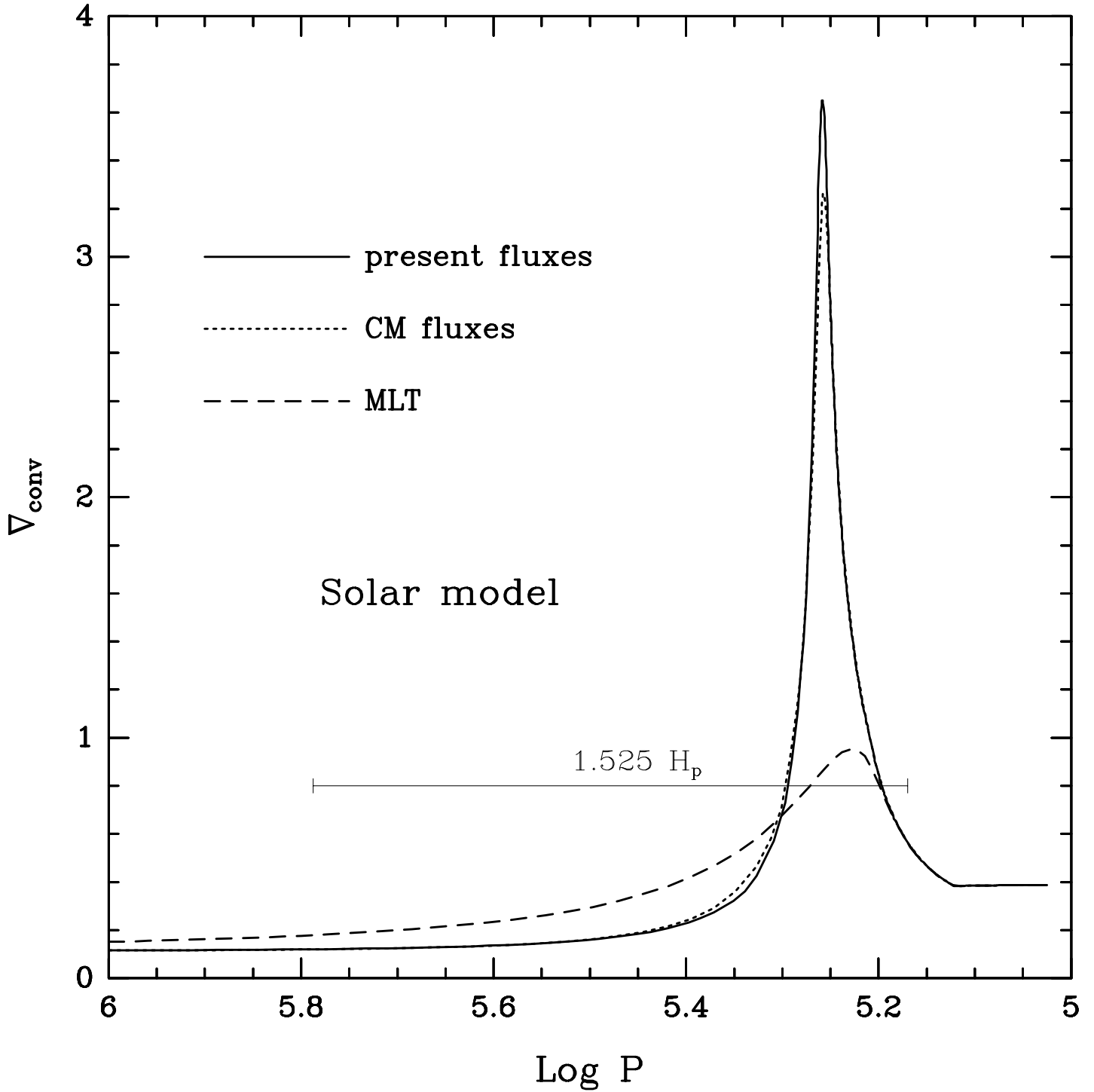}
\vskip - 2 truecm
\caption{The dimensionless temperature gradient, $d\,{\rm 
log}\,T/d\,{\rm log}\,P$, versus pressure in the
upper convective layer of the sun. The solid line corresponds to the  
present model. The  dotted line corresponds to the CM model. The MLT 
(with $\alpha=1.55$)
yields quite different results represented by the dashed line.}
\end{figure}

\begin{figure}
\vskip - 5 truecm
\plotone{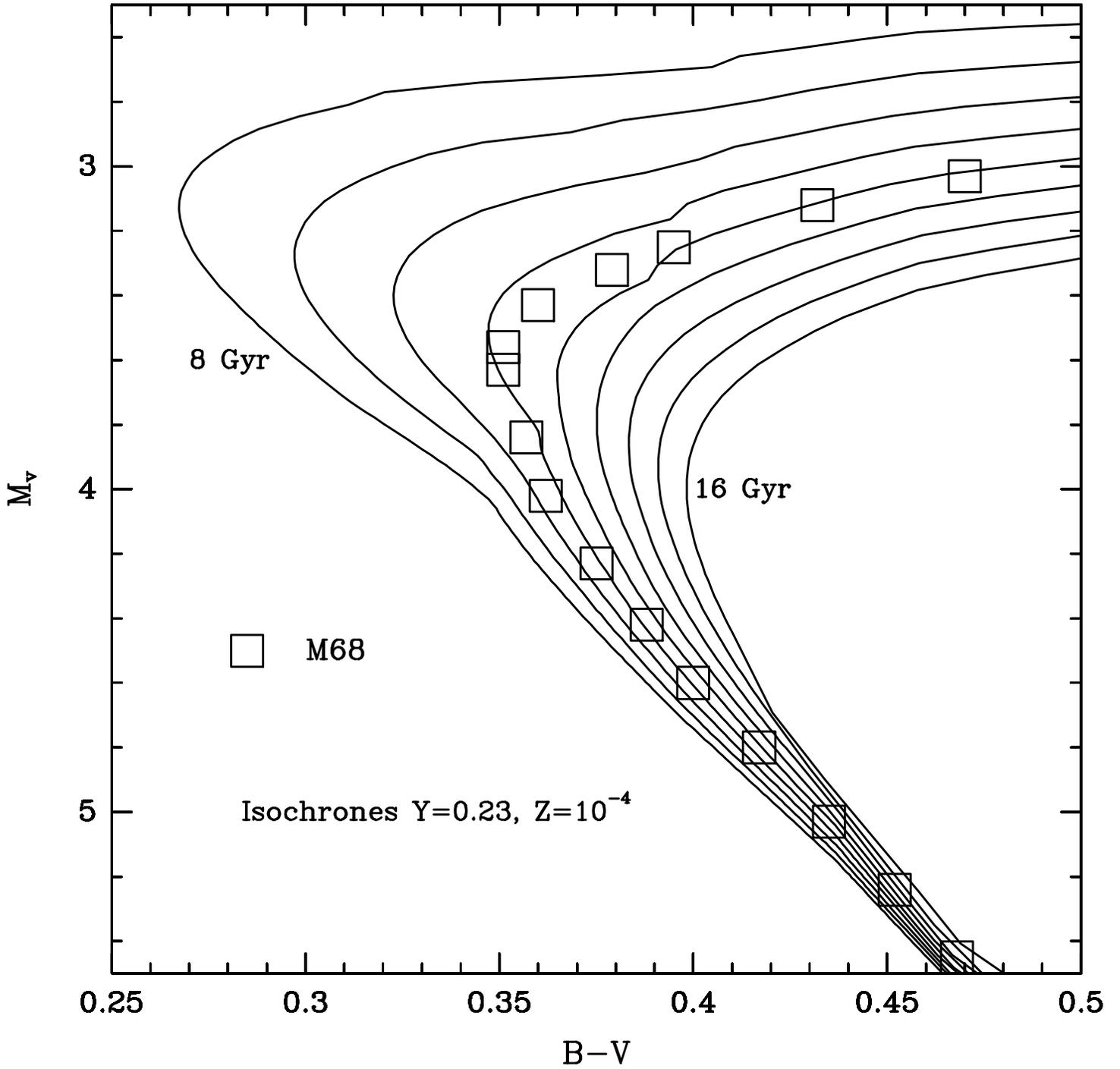}
\vskip  1 truecm
\caption{Isochrones in the HR diagram computed with the 
present model for 
an extreme Pop II chemical composition ($Y=0.23$ and $Z=10^{-4}$). The 
squares mark the
fiducial Turn Off region for the very metal-poor globular cluster M68.}
\end{figure}

\end{document}